\newif\ifcomments  
\commentstrue

\documentclass[a4paper,table]{article}
\usepackage{INTERSPEECH2022}
\usepackage[dvipsnames]{xcolor}
\usepackage{highlight2}
\usepackage{comment}
\usepackage{hyperref}
\usepackage{url}
\usepackage{graphicx}
\usepackage{enumitem}

\title{E2E Segmenter: Joint Segmenting and Decoding for Long-Form ASR}
\name{W. Ronny Huang, Shuo-yiin Chang, David Rybach, Rohit Prabhavalkar, \\Tara N. Sainath, Cyril Allauzen, Cal Peyser, Zhiyun Lu}

\address{Google Research, USA}
\email{\{wrh, shuoyiin, rybach, prabhavalkar, tsainath, allauzen, cpeyser, zhiyunlu\}@google.com}

\begin{document}
\maketitle
\begin{abstract}
Improving the performance of end-to-end ASR models on long utterances ranging from minutes to hours is an ongoing challenge in speech recognition.
A common solution is to segment the audio in advance using a separate voice activity detector (VAD) that decides segment boundary locations based purely on acoustic speech/non-speech information. VAD segmenters, however, may be sub-optimal for real-world speech where, e.g., a complete sentence that should be taken as a whole may contain hesitations in the middle (``set an alarm for... 5 o'clock'').

We propose to replace the VAD with an end-to-end ASR model capable of predicting segment boundaries in a streaming fashion, 
allowing the segmentation decision to be conditioned not only on better acoustic features but also on semantic features from the decoded text with negligible extra computation.
In experiments on real world long-form audio (YouTube) with lengths of up to 30 minutes long, we demonstrate 8.5\% relative WER improvement and 250 ms reduction in median end-of-segment latency compared to the VAD baseline on a state-of-the-art Conformer RNN-T model.
\end{abstract}
\noindent\textbf{Index Terms}: speech recognition, speech segmentation, decoding algorithms, beam search




\section{Introduction}

Streaming End-to-end (E2E) models for ASR have achieved low word error rates (WERs) for short to medium length utterances of up to a few minutes long \cite{liu2021exploiting,sainath2021cascadedlm}.
However, E2E models have high WERs and suffer from deletion error on \textit{long-form} utterances of tens of minutes to hours long \cite{chiu2021rnn,lu2021input,wang2022vadoi}.
Such utterances are found in tasks like meetings, lectures, and video captions.


A common practice for processing long-form utterances is to first segment the audio upstream with a separate voice activity detector (VAD).
Whenever the VAD detects a long silence, it splits the audio at that location into two segments \cite{ramirez2007voice,yoshimura2020end},
which are then processed independently by the E2E model.
At each segment boundary, the beam search \textit{finalizes} the top hypothesis by discarding all other hypotheses. 
This introduces more diversity into the beam search by occasionally clearing away stale hypotheses and making room for new ones, ultimately improving WER by seeing more potentially correct hypotheses.
Maintaining beam diversity is particularly important for E2E models which are typically decoded with small beams.

Despite its crucial role in segmenting audio and regimenting the beam search,
very little attention has been paid to improving \textit{end-of-segment} prediction task \cite{ali2018innovative,hou2020segment}.
Current segmenters suffer from high latency because the VAD, by design, must wait through a long silence before deciding to segment.
This delays subsequent functions like rescoring \cite{sainath2019two} or prefetching \cite{chang2020prefetch} that must wait for the hypotheses to be finalized.
Improving the latency is important because it can improve user experience by making smart assistants more responsive via faster prefetching,
or by helping dictation or captioning apps reduce the amount of ``flickering'' due to switching between top hypotheses.
Current segmenters also suffer from high segmentation error because the VAD bases its decision purely on the audio and not the decoded text \cite{li2021long}, which can contain semantic clues as to when to segment.
Improving segmentation correctness is important because it can improve WER.
As a motivating example, consider two segmentations of the spoken audio below.
Note the segment boundary is denoted by ``$\texttt{|}$''. 

\begin{center}
    \footnotesize
    Audio: ``Shaq... dunks---game over!'' \\[1mm]
    S1:\;\scalebox{.8}[1.0]{\texttt{\textbf{shaq dunks | game over}}} \\
    S2:\;\scalebox{.8}[1.0]{\texttt{\textbf{shaq | dunks game over}}}
\end{center}


\noindent
An ideal segmenter would give S1 because it separates the speech into semantically consistent chunks.
However, a VAD segmenter would give S2 because the speaker pauses after saying ``Shaq...''.
Importantly, S2's suboptimal segmentation can lead to a word error if ``shack'' was the top hypothesis during finalization.
This is because by the time ``dunks'' is spoken, there is no more opportunity to revise ``shack'' to ``shaq'' due to the finalization.
On the other hand, not segmenting at all would lead to bloating the beam with no diversity in the hypotheses, which could also induce word errors \cite{prabhavalkar2021less}.

A related problem exists for end-of-query (EOQ) prediction, or endpointing, which historically also used audio-based VAD or EOQ detectors \cite{shannon2017improved}.
Recently, WER and latency gains have been achieved by combining endpointing and ASR into a single E2E model that is jointly optimized on both tasks, allowing them to share acoustic and semantic information \cite{maas2018combining,Shuoyiin19,li2020towards,hwang2020end,lu2022endpoint}.

Taking inspiration from the E2E endpointing work above,
we now introduce E2E Segmenter, an E2E model jointly optimized on both end-of-segment detection and ASR tasks. 
A central challenge to end-to-end segmenting is that unlike the end-of-query label which indisputably belongs at the end of the transcript,
there is no ground truth for where end-of-segment labels ought to be---making supervised training difficult.
We address this challenge by proposing a novel end-of-segment annotation scheme based on modeling hesitations and word timings.
To avoid degrading wordpiece prediction, we also introduce a new joint layer in the RNN-T architecture that independently predicts the end-of-segment token while leveraging shared acoustic and semantic features.
Compared to the VAD baseline, E2E segmenter achieves quality improvements of up to 8.5\% WER relative while simultaneously reducing 50th percentile latency by 250 ms on the YouTube captioning task.

\section{Method}
The primary job of the segmenter is to send \textit{segment boundary signals} to the beam search in a streaming fashion.
Upon receiving this signal, the beam search finalizes the top hypothesis, clears the beam, resets the encoder state,
and passes the top-hypothesis decoder state to the new segment.
The decision of \textit{when} to send the segment boundary signal is conventionally made by an upstream VAD model;
but here, the signal is produced by the decoder itself whenever the top hypothesis in the beam search predicts it has reached the end-of-segment with confidence above a threshold.
We now discuss how the E2E model is designed to perform this end-of-segment (\texttt{<eos>}) prediction task.

\subsection{End-of-segment joint layer}
\begin{figure}[]
  \centering
  \includegraphics[width=0.85\linewidth]{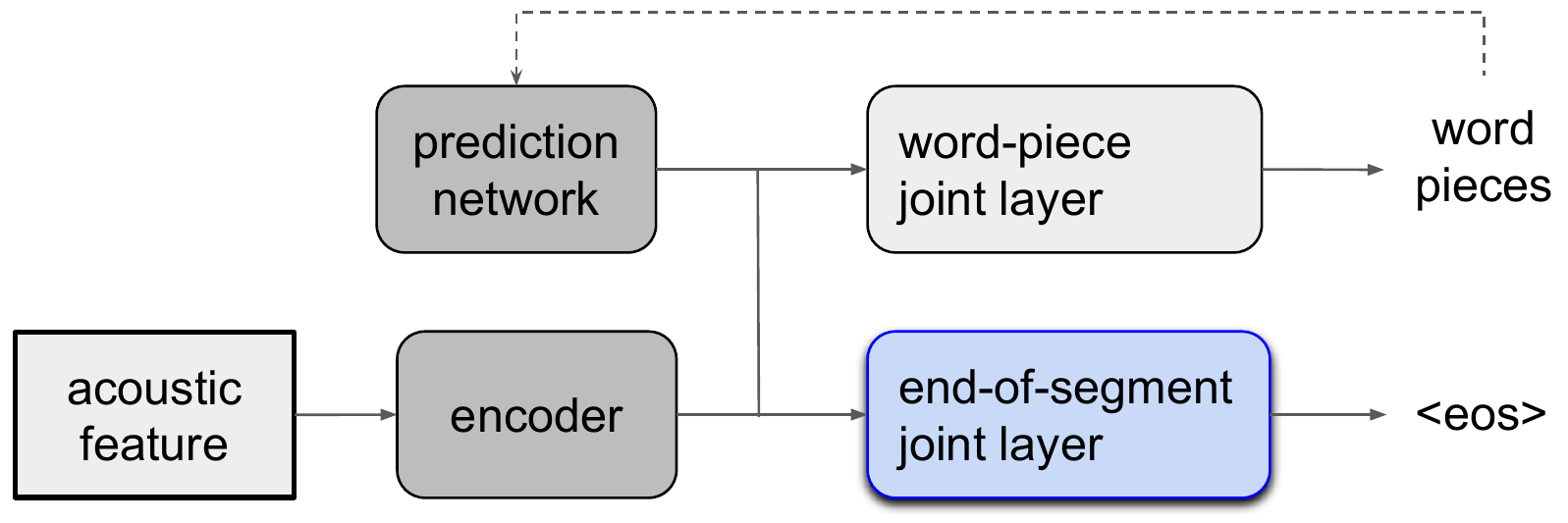}
  \caption{RNN-T with additional joint layer for emitting the end-of-segment posterior.}
  \label{fig:architecture}
\end{figure}

Figure \ref{fig:architecture} illustrates our architecture, which is similar to that in \cite{chang2021turn}.
The original RNN-T wordpiece joint network is a shallow, single layer of the RNN-T model that fuses both acoustic (from the encoder) and linguistic (from the prediction network) sources of information and emits token posteriors.
A natural way of conferring the end-of-segment prediction task to the RNN-T decoder would be assign the joint layer an additional output logit representing \texttt{<eos>},
as is done for endpointing \cite{Shuoyiin19,li2020towards},
but we found in pilot experiments that this interferes with wordpiece decoding and hurts WER.
Instead, to decouple wordpiece prediction from end-of-segment prediction, we add a second joint layer---the end-of-segment joint layer---that emits an \texttt{<eos>} posterior, i.e.

\begin{equation}
    P(\texttt{<eos>}|\textbf{x}_1, \cdots, \textbf{x}_t, y_1, \cdots, y_u)
\end{equation}

\noindent
where $\textbf{x}_i$ is the i-th audio frame and $y_i$ is the i-th decoded token in the beam.
The end-of-segment joint layer is identical in structure to the wordpiece joint layer, containing all wordpieces as logits.
Standard wordpiece training of the RNN-T model with the wordpiece joint layer first occurs;
then the end-of-segment joint layer is initialized with the same weights as the wordpiece joint layer and fine-tuned on the training data with \texttt{<eos>} prediction included.
During inference, the wordpiece joint layer is used for wordpiece prediction while the end-of-segment joint layer is used for end-of-segment prediction.

\subsection{End-of-segment annotation}
While the architecture now allows for emitting the \texttt{<eos>} token,
how can we train the model to emit it at the appropriate time?
What patterns from the audio or text data can be used as supervision for when an \texttt{<eos>} \textit{ought to} occur?
Human annotation is expensive and inconsistent---it is not even clear in principle where best to insert segment boundaries.
Thus, we opt for a heuristic-based, weak supervision approach where \texttt{<eos>} ground truth labels are automatically inserted into the training transcripts based on the rules shown in Table \ref{tab:heuristics}.

\begin{figure}[]
  \centering
  \includegraphics[width=\linewidth]{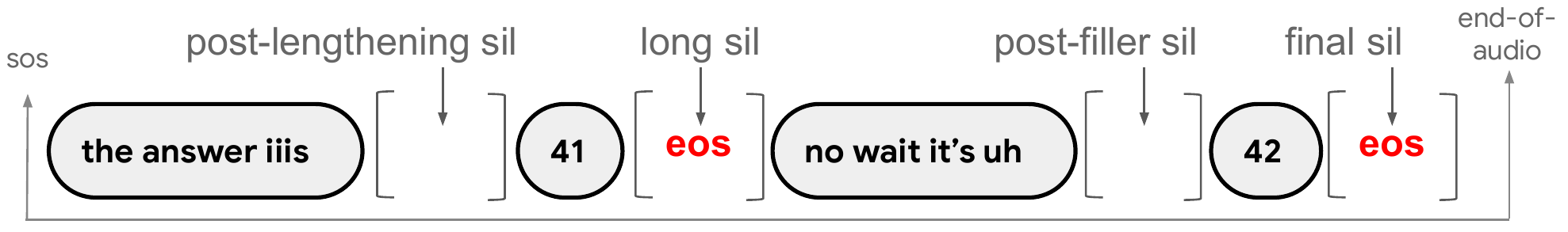}
  \caption{Example of \texttt{<eos>} annotation. ``sil'' = silence.}
  \label{fig:annotation}
\end{figure}

These heuristics include rules for inserting an \texttt{<eos>} when there is a long silence ($\ge$1.2s) or at the end of the utterance.
To eliminate common mis-insertions,
we also specify two exceptions for patterns where the model might otherwise insert \texttt{<eos>},
but are in fact places where the speaker is likely not finished with the sentence.
Refer to Figure \ref{fig:annotation} for an example.
Specifically these include silences following lengthened words (heyyy) or filler words (um) which signal speaker hesitation.
We identify as lengthened words those with a phoneme duration exceeding 5 times the standard deviation;
and we use an in-house model to detect filler words.
Implementing these heuristics required obtaining silence, word, and phoneme timings via running a forced alignment model on all audio-text pairs in the training set.

\begin{table}[]
\centering
\caption{Rules and exceptions for inserting \texttt{<eos>} annotations.}
\label{tab:heuristics}
\resizebox{\columnwidth}{!}{%
\begin{tabular}{llll}
\toprule 
 &
  \begin{tabular}[c]{@{}l@{}}\quad\\ When this happens...\end{tabular} &
  \begin{tabular}[c]{@{}l@{}}\quad\\ It's likely because...\end{tabular} &
  \begin{tabular}[c]{@{}l@{}}Therefore,\\ insert...\end{tabular} \\
\midrule
Rule 1 & Long silence between words  & Speaker finished & \texttt{<eos>} \\[2mm]
Rule 2 & Silence following last word & Speaker finished & \texttt{<eos>} \\
\midrule
\begin{tabular}[c]{@{}l@{}}\quad\\ Exception 1\end{tabular} &
  \begin{tabular}[c]{@{}l@{}}Silence following lengthened\\ word (e.g. heyyy)\end{tabular} &
  \begin{tabular}[c]{@{}l@{}}\quad\\ Speaker not finished\end{tabular} &
  \begin{tabular}[c]{@{}l@{}}\quad\\ nothing\end{tabular} \\[4mm]
\begin{tabular}[c]{@{}l@{}}\quad\\ Exception 2\end{tabular} &
  \begin{tabular}[c]{@{}l@{}}Silence following\\ filler word (e.g. um)\end{tabular} &
  \begin{tabular}[c]{@{}l@{}}\quad\\ Speaker not finished\end{tabular} &
  \begin{tabular}[c]{@{}l@{}}\quad\\ nothing\end{tabular} \\
\bottomrule
\end{tabular}%
}
\end{table}

\subsection{FastEmit training}
Now that the model emits \texttt{<eos>} \textit{correctly}, we wish to make it emit \textit{quickly}.
After all, one of the advantages of E2E segmenting is that it does not need to wait a fixed silence duration before emitting \texttt{<eos>} like the VAD.
Therefore, we train our model with the FastEmit regularization term \cite{yu21fastemit} which encourages each token to be emitted as soon as sufficient context is available.
During inference, the FastEmit-trained model can emit \texttt{<eos>} sooner than the silence duration required to insert that token during the ground truth annotation procedure (Table \ref{tab:heuristics}, Rule 1).

\section{Setup}

\begin{table}[]
\centering
\caption{Length statistics of YouTube testsets.}
\label{tab:yt}
\resizebox{\columnwidth}{!}{%
\begin{tabular}{l|lllll}
\toprule
         &        &             &           & \multicolumn{2}{c}{Percentile} \\
Test set & \# Utt. & Tot words & Tot. length & 50th               & 75th              \\
\midrule
YT\_LONG  & 77     & 207191    & 22.2h       & 14.8m              & 30m               \\
YT\_SHORT & 105    & 84862     & 9.0h        & 6.3m               & 7.4m              \\
\bottomrule
\end{tabular}%
}
\end{table}

\begin{table*}[h!]
\centering
\caption{(a) Main results. (b) End-of-segment threshold ablation study. Naming convention is E2E-\{eos\_threshold\_value\}}.
\vspace{-7pt}
\label{tab:main}
\begin{tabular}{ll|lllll|lllll}
\toprule
                                    & & \multicolumn{5}{c}{YT\_LONG}                               & \multicolumn{5}{c}{YT\_SHORT} \\
    & Segmenter      & WER   & EOS50 & EOS75 & \# Seg. & \# State & WER & EOS50 & EOS75 & \# Seg. & \# State \\
\midrule
(a) & B1: Fixed-10s  & \cg{20.05}{17.05}{20.05} & -   & -   & \cg{99.8}{51.3}{99.8}   & \cg{5496}{5496}{8396}   & \cg{13.63}{10.49}{13.63} & -     & -     & \cg{33.9}{17.5}{33.9}  & \cg{3965}{3965}{5689}      \\
    & B2: Fixed-20s  & \cg{18.22}{17.05}{20.05} & -   & -   & \cg{51.3}{51.3}{99.8}   & \cg{5753}{5496}{8396}   & \cg{11.57}{10.49}{13.63} & -     & -     & \cg{17.5}{17.5}{33.9}  & \cg{4048}{3965}{5689}      \\
    & B3: VAD        & \cg{18.16}{17.05}{20.05} & \cg{260}{130}{260} & \cg{490}{353}{490} & \cg{95.2}{51.3}{99.8}   & \cg{8396}{5496}{8396}   & \cg{11.46}{10.49}{13.63} & \cg{460}{210}{460}   & \cg{660}{365}{660}   & \cg{28.3}{17.5}{33.9}  & \cg{5689}{3965}{5689}      \\
    & E1: E2E (best) & \bg{17.05}{17.05}{20.05} & \bg{130}{130}{260} & \bg{353}{353}{490} & \cg{56.2}{51.3}{99.8}   & \cg{8098}{5496}{8396}   & \bg{10.49}{10.49}{13.63} & \bg{210}{210}{460}   & \bg{365}{365}{660}   & \cg{18.2}{17.5}{33.9}  & \cg{5672}{3965}{5689}      \\
    & $\Rightarrow$\;\; E1 vs. B3 & \textcolor{red}{-6.1\%} & \textcolor{red}{-130} & \textcolor{red}{-137} &  &  & \textcolor{red}{-8.5\%} & \textcolor{red}{-250} & \textcolor{red}{-265} &  &       \\
\midrule                                                                                                       
(b) & E2: E2E-0.0      & \cg{17.60}{17.05}{17.76} & \cg{180}{75}{220} & \cg{550}{210}{550}  & \cg{16.4 }{16.4}{477.6}  & \cg{7244}{5410}{8513}   & \cg{11.10}{10.49}{11.10} & \cg{510}{140}{510}	& \cg{855}{230}{855}  & \cg{5.8  }{5.8}{119.1} & \cg{5416}{3975}{5771}      \\
    & E3: E2E-0.5      & \cg{17.38}{17.05}{17.76} & \cg{180}{75}{220} & \cg{550}{210}{550}  & \cg{18.7 }{16.4}{477.6}  & \cg{7958}{5410}{8513}   & \cg{10.91}{10.49}{11.10} & \cg{460}{140}{510}	& \cg{800}{230}{855}  & \cg{6.8  }{5.8}{119.1} & \cg{5705}{3975}{5771}      \\
    & E4: E2E-1.0      & \cg{17.19}{17.05}{17.76} & \cg{220}{75}{220} & \cg{535}{210}{550}  & \cg{26.3 }{16.4}{477.6}  & \cg{8513}{5410}{8513}   & \cg{10.68}{10.49}{11.10} & \cg{390}{140}{510}	& \cg{745}{230}{855}  & \cg{9.6  }{5.8}{119.1} & \cg{5771}{3975}{5771}      \\
    & E5: E2E-1.5      & \cg{17.09}{17.05}{17.76} & \cg{195}{75}{220} & \cg{445}{210}{550}  & \cg{38.6 }{16.4}{477.6}  & \cg{8366}{5410}{8513}   & \cg{10.63}{10.49}{11.10} & \cg{280}{140}{510}	& \cg{395}{230}{855}  & \cg{13.3 }{5.8}{119.1} & \cg{5700}{3975}{5771}      \\
    & E6: E2E-2.0      & \bg{17.05}{17.05}{17.76} & \cg{130}{75}{220} & \cg{353}{210}{550}  & \cg{56.2 }{16.4}{477.6}  & \cg{8098}{5410}{8513}   & \bg{10.49}{10.49}{11.10} & \cg{210}{140}{510}	& \cg{365}{230}{855}  & \cg{18.2 }{5.8}{119.1} & \cg{5672}{3975}{5771}      \\
    & E7: E2E-2.5      & \cg{17.08}{17.05}{17.76} & \cg{100}{75}{220} & \cg{355}{210}{550}  & \cg{80.7 }{16.4}{477.6}  & \cg{7917}{5410}{8513}   & \bg{10.49}{10.49}{11.10} & \cg{200}{140}{510}	& \cg{345}{230}{855}  & \cg{25.7 }{5.8}{119.1} & \cg{5371}{3975}{5771}      \\
    & E8: E2E-3.0      & \cg{17.06}{17.05}{17.76} & \cg{100}{75}{220} & \cg{415}{210}{550}  & \cg{116.3}{16.4}{477.6}  & \cg{7362}{5410}{8513}   & \cg{10.58}{10.49}{11.10} & \cg{180}{140}{510}	& \bg{230}{230}{855}  & \cg{35.6 }{5.8}{119.1} & \cg{5039}{3975}{5771}      \\
    & E9: E2E-3.5      & \cg{17.22}{17.05}{17.76} & \cg{90 }{75}{220} & \cg{280}{210}{550}  & \cg{178.5}{16.4}{477.6}  & \cg{6926}{5410}{8513}   & \cg{10.56}{10.49}{11.10} & \cg{180}{140}{510}	& \cg{450}{230}{855}  & \cg{49.8 }{5.8}{119.1} & \cg{4790}{3975}{5771}      \\
    & E10: E2E-4.0     & \cg{17.48}{17.05}{17.76} & \cg{75 }{75}{220} & \bg{210}{210}{550}  & \cg{300.0}{16.4}{477.6}  & \cg{6237}{5410}{8513}   & \cg{10.58}{10.49}{11.10} & \cg{180}{140}{510}	& \cg{380}{230}{855}  & \cg{74.7 }{5.8}{119.1} & \cg{4569}{3975}{5771}      \\
    & E11: E2E-4.5     & \cg{17.76}{17.05}{17.76} & \bg{90 }{75}{220} & \cg{255}{210}{550}  & \cg{477.6}{16.4}{477.6}  & \cg{5410}{5410}{8513}   & \cg{10.78}{10.49}{11.10} & \bg{140}{140}{510}	& \cg{280}{230}{855}  & \cg{119.1}{5.8}{119.1} & \cg{3975}{3975}{5771}      \\
\bottomrule
\end{tabular}%
\end{table*}

\subsection{Dataset}
\label{sec:datasetup}
YouTube videos cover many domains (TV shows, sports, conversations, etc.) and are often be very long \cite{narayanan2019recognizing},
making YouTube captioning an ideal task for our long-form study.
Thus, we evaluate on two standard YouTube testsets used in \cite{Soltau2017,chiu2019comparison,chiu2021rnn}:
YT\_LONG is sampled from YouTube video-on-demand and YT\_SHORT is sampled from Google Preferred channels on YouTube.
Table \ref{tab:yt} shows their length statistics.

The training set, identical to that in \cite{sainath2020streaming}, is a sample of Google traffic from multiple domains such as voice search, farfield, telephony, and YouTube, making up about 300M utterances with 400k hours of audio.
All utterances are anonymized and hand-transcribed, with the exception of YouTube being semi-supervised \cite{liao2013large}.
Note the YouTube utterances used for training are cut into small chunks no more than about 20 seconds long.
The data is diversified via multi-style training \cite{kim2017mtr}, random down-sampling from 16 to 8 kHz \cite{li2012improving}, and SpecAug \cite{Park2019}.

\subsection{Model}
Our RNN-T model is similar to the first-pass network of \cite{sainath2021cascadedlm}.
The encoder is a streaming 12-layer, 512 dimensional Conformer encoder with causal convolution kernels of size 15 and 8 left-context self-attention heads.
The decoder consists of a stateless prediction network \cite{Rami21} with output dimension 640.
The joint layers (both wordpiece and end-of-segment) are single layers which input the concatenation of encoder and prediction network features.
In total, the model has 140M parameters, of which less than 1M are due to the additional end-of-segment joint layer.
The model emits 4096 wordpieces, with the blank token factored out with HAT factorization \cite{variani2020hybrid}.
Model training minimizes the RNN-T and MWER loss \cite{prabhavalkar2018minimum}.
We also add the FastEmit regularization term \cite{yu21fastemit} with a weight of 5e-3.
The optimizer was Adam with $\beta_1=0.9$ and $\beta_2=0.999$.
A transformer learning rate schedule \cite{Vaswani17} with peak learning rate of 1.8e-3 and 32k steps of warm-up is used,
along with exponential-moving-average-stabilized gradient updates.
All models are implemented in Lingvo \cite{shen2019lingvo} and trained on 64 TPU chips with a global batch size of 4096 for 500k steps. 

\subsection{Beam search}
We use a frame-synchronous beam search with a beam size of 8 and a pruning threshold of 5;
i.e. partial hypotheses with negative log posterior exceeding that of the top hypothesis by 5 are removed.
At each frame, we apply a breadth-first search for possible expansions similar to \cite{tripathi2019monotonic}, ignoring any expansion with a negative log posterior of 5 or greater, and limiting the search depth to 10 expansions. 
The production streaming client we run on has a maximum segment duration of 65 seconds before it forces a finalization.

\subsection{Voice activity detector}
Our pipeline contains a lightweight voice activity detector \cite{zazo2016feature} upstream of the E2E model that classifies each frame as silence or speech in a streaming fashion.
Whenever it detects 0.2 seconds of continued silence,
it sends a segment boundary signal forcing the beam search to reset encoder state and discard all except the top hypothesis.
VAD-based segment finalization is turned on only in our baselines; it is turned off for all E2E segmenter experiments.

\section{Results}
\label{sec:results}
In Table \ref{tab:main}, we run the ASR pipeline with different segmenters on YT\_LONG and YT\_SHORT.
Other than the segmenter, all aspects of the ASR pipeline are identical.
We track the following metrics for each experiment:
\begin{itemize}[leftmargin=*]
\item \textbf{WER}:
Word error rate---measure of overall ASR quality.
\item \textbf{EOS50}, \textbf{EOS75}:
End-of-segment latency in milliseconds,
i.e., how long after speaking does the transcription get finalized.
Since the only segment boundary that can be considered ground truth is the one at the end of the utterance,
we measure the time difference from the end of the last word (whose timing is determined by forced alignment) to the last segment boundary, averaged across utterances.
We report the 50th and 75th percentile EOS latencies.
Anomalous latencies below -0.5s or exceeding 2s are left out of the percentile calculation.
\item \textbf{\# Segment}:
Average number of segments for each utterance.
\item \textbf{\# State}:
Average number of model states in the beam search for each utterance.
This metric, used also in \cite{prabhavalkar2021less},
is equivalent to the number of joint network forward passes and is thus a measure of the beam search efficiency.
\end{itemize}

\subsection{Main results}
In Table \ref{tab:main}a, we first show that the quality of the segmentation matters by presenting two baseline fixed-interval segmenters, B1 and B2,
and comparing it to the VAD segmenter, B3.
The VAD segmentation is determined by silences and achieves better WER than the fixed-length segmenters which do not depend on any features. 

Next we pick E1, our best E2E segmenter (operating point determined in \S\ref{sec:eosablation}), and compare it against B3, the VAD segmenter.
E1 outperforms B3 by 6.1\% WER relative on YT\_LONG and 8.5\% on YT\_SHORT,
highlighting the segmenter's ability to improve overall quality.
E1 also finalizes the segments faster than the VAD by 130/137 ms on YT\_LONG, measured at 50th/75th percentiles, and by 250/265 ms on YT\_SHORT.
These improvements are within perceivable range for user experience.

E1 also achieves a slightly better beam search efficiency (lower number of states), which may be due to the fact that its hypotheses are more stable, obviating the need for many joint expansions. 
The number of segments in the VAD and E2E are similar to those in the two fixed-length segmenters (B1-B2), which isolates out the fact that segmentation correctness matters as opposed to the number of segments.

\subsection{\texttt{<eos>} threshold ablation study}
\label{sec:eosablation}
Table \ref{tab:main}b shows an ablation study on the \texttt{<eos>} threshold.
When the \texttt{<eos>} negative log-posterior from the model falls below the \texttt{<eos>} threshold, the segment is finalized.
Higher thresholds finalize more aggressively, leading to lower latency and more segments,
but at the cost of more segmentation errors, e.g., finalizing in the middle of a sentence.
Conversely, lower thresholds may not finalize as often as is needed to maintain beam diversity.
The sweet spot for WER occurs---for both test sets---at a threshold of 2.0, and we pick that as our operating point.


\subsection{Utterance length dependence}
\begin{figure}[]
  \centering
  \includegraphics[width=\linewidth]{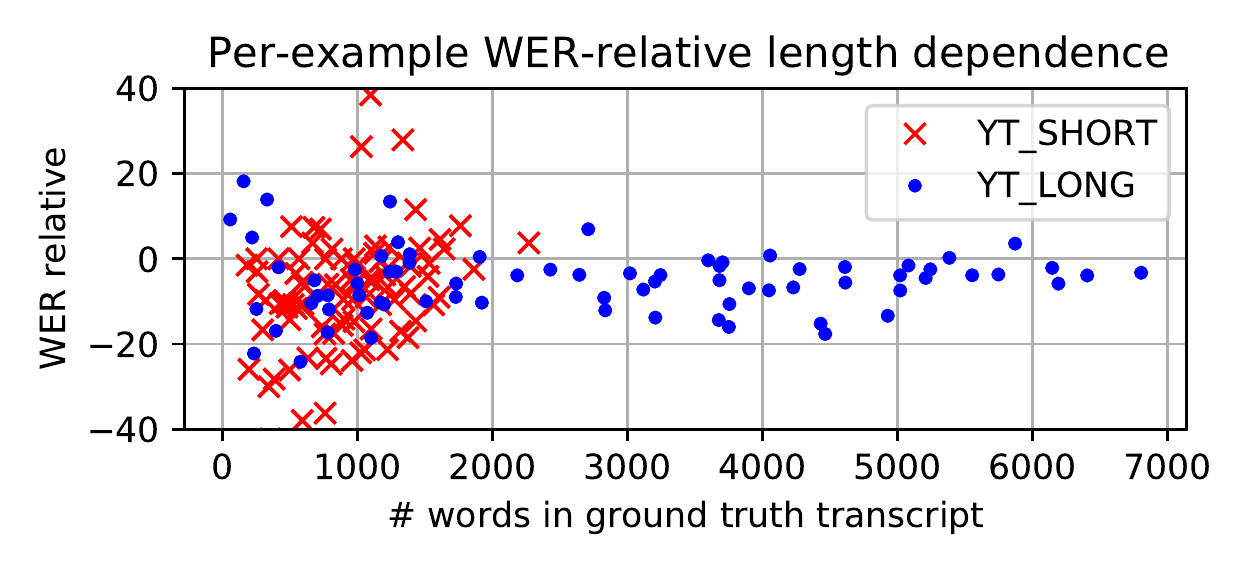}
  \vspace{-20pt}
  \caption{Per-example WER-relative of E2E (E1) to VAD (B3) segmenters versus utterance length. Lower is better.}
  \label{fig:perexample}
\end{figure}

In Figure \ref{fig:perexample}, we evaluate the \textit{per-example} WER-relative between E1 and B3 as a function of utterance length.
We define utterance length here as the number of words in the ground truth transcript rather than the audio duration (though they are correlated)
because it corresponds more closely to the beam search lattice length.
This allows us to analyze whether our WER gains are limited to long-form utterances.
Surprisingly, WER-relative is rather invariant of utterance length for both test sets, even for utterances with a few hundred words (a few minutes).
This suggests that E2E segmentation can be applied more widely to medium-form utterances as well.

\subsection{Results with frame filtering}
\label{sec:framefilt}
In Table \ref{tab:framefilt}, we evaluate the VAD and E2E segmenters with frame filtering turned \textit{on}.
Like VAD-based finalization, frame filtering starts when the VAD detects 0.2 seconds of silence,
jettisoning forthcoming frames until speech is detected again.
This is a practical measure to save computation for on-device deployments,
because it prevents silence frames from being unnecessarily processed by the expensive E2E model.
Segmenting and frame filtering are conventionally tightly coupled;
when the VAD decides to segment, it simultaneously kicks off frame filtering,
ensuring that the segmentation decision has access to all the audio frames.
Replacing the segmenting with an E2E model requires an assessment of how it interacts with the VAD-controlled frame filtering, since segmenting may now happen before or after frame filtering begins.

A first observation is that frame filtering increases absolute WER by around 2\% compared to no frame filtering due to the reduced acoustic context (see B4 vs. B3 and E12 vs. E1). 
However E2E still prevails over VAD by about 3.1\% WER relative (E12 vs. B4) and 120 ms EOS50 latency.
It also achieves better beam search efficiency (lower number of states), which is aligned with frame filtering's goal of reducing computational load.
Compared to no frame filtering (E1), E12's number of segments is decreased from 56.1 to 28.1.
This is because the model, though trained with FastEmit, still needs to see some silence in order to confidently predict end-of-segment, and overly aggressive frame filtering prevents that silence from being seen.

The frame filtering can be gradually reduced by increasing its \textit{margin}, or the additional silence time beyond 0.2 seconds the VAD must detect before initializing frame filtering.
As the margin is increased in E13-E20,
the WER converges towards its value without frame filtering (17.05\%), at the cost of slightly increasing computation.
The EOS50 latency is also reduced; investigating the cause of this is a point of future work.
This table suggests that a trade-off between quality and beam search efficiency must be made in resource-constrained situations with frame filtering.

\begin{table}[t]
\centering
\caption{Segmenting with frame filtering for YT\_LONG. YT\_SHORT results are similar and not displayed for brevity. Naming convention is \{segmenter\}-\{margin\_length\}}
\vspace{-5pt}
\label{tab:framefilt}
\begin{tabular}{l|llll}
\toprule
 & \multicolumn{4}{c}{YT\_LONG} \\
Segmenter & WER & EOS50 & \# Seg & \# State \\
\midrule
B4: VAD-0s     & \cg{20.59}{17.12}{20.59} & \cg{260}{80}{260} & \cg{94.4}{28}{95} & \cg{7108}{6743}{7228} \\
E12: E2E-0s    & \bg{19.94}{17.12}{20.59} & \bg{140}{80}{260} & \cg{28.1}{28}{95} & \bg{6799}{6743}{7228} \\
$\Rightarrow$ E12 vs. B4 & \textcolor{red}{-3.1\%} & \textcolor{red}{-120} & & \\
\midrule
E13: E2E-1s    & \cg{19.55}{17.12}{20.59} & \cg{85 }{80}{260} & \cg{33.0}{28}{95} & \cg{6834}{6743}{7228} \\
E14: E2E-2s    & \cg{19.21}{17.12}{20.59} & \cg{85 }{80}{260} & \cg{41.9}{28}{95} & \cg{6776}{6743}{7228} \\
E15: E2E-4s    & \cg{18.71}{17.12}{20.59} & \bg{80 }{80}{260} & \cg{48.9}{28}{95} & \bg{6743}{6743}{7228} \\
E16: E2E-8s    & \cg{18.05}{17.12}{20.59} & \cg{105}{80}{260} & \cg{53.1}{28}{95} & \cg{6981}{6743}{7228} \\
E17: E2E-16s   & \cg{17.52}{17.12}{20.59} & \cg{90 }{80}{260} & \cg{56.2}{28}{95} & \cg{7050}{6743}{7228} \\
E18: E2E-32s   & \cg{17.18}{17.12}{20.59} & \cg{90 }{80}{260} & \cg{56.8}{28}{95} & \cg{7023}{6743}{7228} \\
E19: E2E-64s   & \cg{17.12}{17.12}{20.59} & \cg{90 }{80}{260} & \cg{56.3}{28}{95} & \cg{7228}{6743}{7228} \\
E20: E2E-128s  & \bg{17.12}{17.12}{20.59} & \cg{90 }{80}{260} & \cg{56.8}{28}{95} & \cg{7058}{6743}{7228} \\
\bottomrule
\end{tabular}%
\end{table}

\section{Conclusion}
Our work presents a way to improve streaming long-form audio decoding by replacing the VAD-based segmenter with an E2E model.
We proposed an E2E architecture that predicts segment boundaries
and provided an automatic end-of-segment data annotation strategy required for learning that task in an end-to-end fashion.
Our results demonstrate significant WER and end-of-segment latency improvements compared to a VAD baseline on a long-form YouTube captioning task.


\bibliographystyle{IEEEtran}
\bibliography{bibliography}

\end{document}